\newif\iffigs\figstrue
\documentclass[11pt]{report}
\iffigs
   \input{epsf}
\else
\fi
\newsavebox{\uuunit}
\sbox{\uuunit}
    {\setlength{\unitlength}{0.825em}
     \begin{picture}(0.6,0.7)
        \thinlines
        \put(0,0){\line(1,0){0.5}}
        \put(0.15,0){\line(0,1){0.7}}
        \put(0.35,0){\line(0,1){0.8}}
       \multiput(0.3,0.8)(-0.04,-0.02){12}{\rule{0.5pt}{0.5pt}}
     \end {picture}}

\def\IP{\relax{\rm I\kern-.18em P}}

\begin{document}
%
\font\cmss=cmss10 \font\cmsss=cmss10 at 7pt
\def\twomat#1#2#3#4{\left(\matrix{#1 & #2 \cr #3 & #4}\right)}
\def\inbar{\vrule height1.5ex width.4pt depth0pt}
\def\IC{\relax\,\hbox{$\inbar\kern-.3em{\rm C}$}}
\def\IG{\relax\,\hbox{$\inbar\kern-.3em{\rm G}$}}
\def\IB{\relax{\rm I\kern-.18em B}}
\def\ID{\relax{\rm I\kern-.18em D}}
\def\IL{\relax{\rm I\kern-.18em L}}
\def\IF{\relax{\rm I\kern-.18em F}}
\def\IH{\relax{\rm I\kern-.18em H}}
\def\II{\relax{\rm I\kern-.17em I}}
\def\IN{\relax{\rm I\kern-.18em N}}
\def\IP{\relax{\rm I\kern-.18em P}}
\def\IQ{\relax\,\hbox{$\inbar\kern-.3em{\rm Q}$}}
\def\bfzero{\relax\,\hbox{$\inbar\kern-.3em{\rm 0}$}}
\def\IK{\relax{\rm I\kern-.18em K}}
\def\IG{\relax\,\hbox{$\inbar\kern-.3em{\rm G}$}}
 \font\cmss=cmss10 \font\cmsss=cmss10 at 7pt
\def\IR{\relax{\rm I\kern-.18em R}}
\def\ZZ{\relax\ifmmode\mathchoice
{\hbox{\cmss Z\kern-.4em Z}}{\hbox{\cmss Z\kern-.4em Z}}
{\lower.9pt\hbox{\cmsss Z\kern-.4em Z}}
{\lower1.2pt\hbox{\cmsss Z\kern-.4em Z}}\else{\cmss Z\kern-.4em
Z}\fi}
\def\bfone{\relax{\rm 1\kern-.35em 1}}
\def\dop{{\rm d}\hskip -1pt}
\def\real{{\rm Re}\hskip 1pt}
\def\trace{{\rm Tr}\hskip 1pt}
\def\ii{{\rm i}}
\def\diag{{\rm diag}}
\def\sch#1#2{\{#1;#2\}}
\def\bfone{\relax{\rm 1\kern-.35em 1}}
\font\cmss=cmss10 \font\cmsss=cmss10 at 7pt
\def\a{\alpha} \def\b{\beta} \def\d{\delta}
\def\e{\epsilon} \def\c{\gamma}
\def\G{\Gamma} \def\l{\lambda}
\def\L{\Lambda} \def\s{\sigma}
\def\cA{{\cal A}} \def\cB{{\cal B}}
\def\cC{{\cal C}} \def\cD{{\cal D}}
\def\cF{{\cal F}} \def\cG{{\cal G}}
\def\cH{{\cal H}} \def\cI{{\cal I}}
\def\cJ{{\cal J}} \def\cK{{\cal K}}
\def\cL{{\cal L}} \def\cM{{\cal M}}
\def\cN{{\cal N}} \def\cO{{\cal O}}
\def\cP{{\cal P}} \def\cQ{{\cal Q}}
\def\cR{{\cal R}} \def\cV{{\cal V}}\def\cW{{\cal W}}
\newcommand{\be}{\begin{equation}}
\newcommand{\ee}{\end{equation}}
\newcommand{\bea}{\begin{eqnarray}}
\newcommand{\eea}{\end{eqnarray}}
\let\la=\label \let\ci=\cite \let\re=\ref
%
%
%
\def\crr{\crcr\noalign{\vskip {8.3333pt}}}
\def\tilde{\widetilde}
\def\bar{\overline}
\def\us#1{\underline{#1}}
\let\shat=\hat
\def\hat{\widehat}
\def\hyp{\vrule height 2.3pt width 2.5pt depth -1.5pt}
\def\square{\mbox{.08}{.08}}
\def\Coeff#1#2{{#1\over #2}}
\def\Coe#1.#2.{{#1\over #2}}
\def\coeff#1#2{\relax{\textstyle {#1 \over #2}}\displaystyle}
\def\coe#1.#2.{\relax{\textstyle {#1 \over #2}}\displaystyle}
\def\half{{1 \over 2}}
\def\shalf{\relax{\textstyle {1 \over 2}}\displaystyle}
\def\dag#1{#1\!\!\!/\,\,\,}
\def\to{\rightarrow}
\def\notin{\hbox{{$\in$}\kern-.51em\hbox{/}}}
\def\shdot{\!\cdot\!}
\def\ket#1{\,\big|\,#1\,\big>\,}
\def\bra#1{\,\big<\,#1\,\big|\,}
\def\equaltop#1{\mathrel{\mathop=^{#1}}}
\def\Trbel#1{\mathop{{\rm Tr}}_{#1}}
\def\inserteq#1{\noalign{\vskip-.2truecm\hbox{#1\hfil}
\vskip-.2cm}}
\def\attac#1{\Bigl\vert
{\phantom{X}\atop{{\rm\scriptstyle #1}}\phantom{X}}}
\def\exx#1{e^{{\displaystyle #1}}}
\def\del{\partial}
\def\delbar{\bar\partial}
\def\nex#1{$N\!=\!#1$}
\def\dex#1{$d\!=\!#1$}
\def\cex#1{$c\!=\!#1$}
\def\eg{{\it e.g.}} \def\ie{{\it i.e.}}
\def\IE{\relax{{\rm I\kern-.18em E}}}
\def\cE{{\cal E}}
\def\rt{{\cR^{(3)}}}
\def\IGam{\relax{{\rm I}\kern-.18em \Gamma}}
\def\IGa{\IA}
\def\ii{{\rm i}}
\begin{titlepage}
\begin{center}
{\Large \bf Branes in Anti de Sitter Space--Time. $^*$ }\\
\vfill
{ M. Trigiante   } \\
\vfill
{\small
 Department of Physics, University of Wales Swansea, Singleton Park,\\
Swansea SA2 8PP, United Kingdom\\
\vspace{6pt}
}
\end{center}
\vfill
\begin{center}
{\bf Abstract}
\end{center}
{\small An intense study of the  relationship between certain quantum theories of gravity realized on  curved backgrounds 
and suitable gauge theories, has been originated by a remarkable conjecture put forward by Maldacena almost one year ago.
Among the possible curved vacua of  superstring or M--theory, spaces having the form of an Anti--de Sitter space--time times a 
compact Einstein manifold, have been playing a special role in this correspondence, since the quantum theory realized on them,
in the original formulation of the conjecture, was identified with the effective superconformal theory on the world volume 
of parallel $p$--branes set on the boundary of such a space (holography). An important step in order to verify such a conjecture and eventually
 generalize it, consists in a precise definition of the objects entering both sides of the holographic correspondence. In 
the most general case indeed it turns out that important features of the field theory on the boundary of the curved background,
identified with the quantum theory of gravity in the bulk, are encoded in the dynamics of the coinciding parallel $p$--branes set on the boundary
of the same space.
The study of $p$--brane dynamics in curved  space--times which are vacua of superstring of M--theory,
 turns out therefore to be a relevant issue in order to verify the existence of the holographic correspondence. 
In the present paper, besides providing a hopefully elementary introduction to Maldacena's duality,
 I shall deal in a  tentatively self contained  way with a particular aspect of the problem of
 $p$-brane dynamics  in Anti--de Sitter space--time,  discussing some recent results. 
}
\vspace{2mm} \vfill \hrule width 3.cm
{\footnotesize Proceeding of a talk given at the XIII Congress of General Relativity  (SIGRAV), Bari 20--27 September 1998.}
\end{titlepage}
\eject
\section{Introduction}
Superstring theory has been regarded for a long time as the most promising quantum theory of gravity (see \cite{wit,pol}).
Indeed it naturally includes the graviton among its massless states. Nevertheless 
it has intrinsic limits deriving from its own definition. As it is well known, this theory 
describes one dimensional extended objects (closed or open strings) vibrating in a ten dimensional Minkowsky space--time,
whose spectrum of vibrational modes are expected to correspond to the observed fields in nature.
From a mathematical point of view the theory is a $\sigma$--model defined on the $1+1$ world sheet of a  string propagating in a $10$--dimensional target space. A dimensionful parameter characterizing the string and defining the 
scale of its vibrational levels is the string length $\ell_s$, whose square will be denoted by $\alpha^\prime$.\par
 Consistency of the theory requires its local invariance with respect to diffeomorphism and conformal transformations on the world sheet coordinates and
 metric. 
Among the massless fields in the spectrum of the theory there is a scalar $\phi$ (the dilaton)
and a spin--$2$ field $G_{\mu \nu}$ which is identified with the graviton field. One of the vacuum solutions of the theory is the one on which 
the symmetric tensor gives the Minkowsky metric on the target space--time ($\langle G_{\mu \nu}\rangle=\eta_{\mu\nu}$)    
and $\langle \phi \rangle =\phi_0$. Superstring theory is quantized on this vacuum, perturbatively with respect to 
an effective coupling constant $g=Exp(\phi_0)$.  Versions of the theory on certain other vacua of the form $M_d\times K$
representing compactifications of the ten dimensional target space to a lower dimensional Minkowsky space $M_d$ ($d\le 10$) times a suitable 
compact 
Ricci--flat manifold $K$ have been constructed as well. The low energy limit of these theories is described by 
a {\it $d$--dimensional supergravity}. 
On the other hand
a second quantized formulation of superstring theory on a curved  background characterized by a non Ricci--flat
 solution of the low--energy effective supergravity theory, is not known so 
far (except for very special cases) and
represents in general a complicate problem (\cite{ktsetse}). As we shall see, one of the potential achievements of the duality recently conjectured by Maldacena (see \cite{malda},\cite{duff} and references therein), is to shed light on the quantum spectrum of superstring theory realized 
on a particular kind of non Ricci--flat vacuum, namely of the form $AdS\times K$, where $AdS$ stands for a $d$--dimensional Anti de Sitter
space--time and $K$ is a suitable compact $(10-d)$--dimensional Einstein space. These vacua correspond to the geometry
of some regions of space--time in presence of extended objects which are believed to belong to the non perturbative spectrum of superstring theories and therefore may be regarded as non--perturbative vacua of these theories.  In the remaining part of the present introduction I shall 
try to further formalize these concepts in a pedagogical fashion, preparing a physical scenario in which to set Maldacena's conjecture.\par 
One of the great achievements of the last ten years is the concept of {\it duality}: correspondences 
have been conjectured and in part verified between regimes of various superstring theories realized on different backgrounds of the form $M_d\times K$
( see \cite{duality} and references therein).
The existence of such dualities would allow to consider the known superstring theories as local effective formulations on different backgrounds of
a unique, yet unknown quantum theory living in a higher dimensional space--time. Among the candidates for this larger quantum theory 
there is the so called $M$--theory
in $(10+1)$--dimensions, whose low energy effective field theory is the well known $11$--dimensional supergravity, and $F$--theory in $(10+2)$--dimensions.
Duality, being a mapping between the spectra of two different superstring theories, at the massless level  is realized by means of 
suitable discrete transformations on the background fields, which close a group $G(Z)$. The largest duality which has been conjectured is called 
$U$--duality and the corresponding action on the massless fields $G(Z)$ is believed to be  suitable discrete version of the largest global symmetry group $G$ 
of the field equations and Bianchi identities of the the underlying low energy supergravity \cite{towhull}.  The study of dualities, and in particular of the 
non--perturbative ones, shed some light on the non--perturbative side of superstring theories. For instance it became clear that the full spectrum of these
theories should contain not only $0$ and $1$--dimensional objects (particles and strings), but also $p$--extended objects ($p$--branes) which were not 
present in the perturbative spectrum, but nonetheless corresponded to solutions of the low energy supergravity. Among them, those coupled with a certain kind of background fields, namely the Ramond--Ramond fields, are of particular interest. The main feature of the Ramond--Ramond (R--R) fields
is that they do not couple directly to the fundamental closed 
string. On the other hand they are mixed with the other background fields 
which do couple directly to the closed string (Neveu-Schwarz--Neveu-Schwarz fields) through
the $U$--duality group $G(Z)$. Therefore, if the $U$--duality is meant to be an 
exact symmetry 
of a larger quantum theory, the distinction between R--R and the other fields
is to be considered an artifact of perturbative superstring theory and therefore one expects to find in the non--perturbative spectrum of superstring theory
objects coupled to R--R fields and thus carrying a R--R charge. Since the R--R fields are in general $(p+1)$ forms, they can minimally couple to $p$--extended objects. A particular kind of these extended objects (namely those 
preserving a fraction of the original supersymmetry) received a description in the context of perturbative superstring theory on flat space as Dirichelet surfaces \cite{tasi}, 
that is (roughly speaking) as hypersurfaces on which open strings start and end. The dynamics of these so called $Dp$--branes  is determined by the 
oscillations of the  open strings attached to them. In ten dimensions the quantum world may be therefore figuratively represented as containing closed strings
free to move in the bulk and to interact among themselves and with dynamical hypersurfaces on which open strings are attached.    
In eleven dimensions on the other hand, the conjectured $M$--theory is expected to contain $p$--extended objects as well ($Mp$--branes), which are 
 solutions of the low energy supergravity.\par Going back to the suggestive $10$--dimensional picture of the quantum world previously portrayed , one may ask 
which are the limits of such a representation, i.e. in what physical regime does the framework of superstring theory on Minkowsky background break down?
A $D$--brane is a massive object, which means that it interacts with the gravitational field and  thus deforms the surrounding space--time. From the superstring point of view, being the quantum fluctuation of the background metric associated with  a vibrational $0$--mode of the closed string, this interaction is 
described through the emission of a closed string from the brane in the surrounding flat space--time and the amplitude of such a process, computed 
in the framework of perturbative string theory by means, for instance, of boundary state methods, yields the deformation of the Minkowsky metric due to the presence of the extended object. For a single $D$--brane this deformation is localized around the source in a region $R_1$ whose thickness is of the order of 
the string length $\ell_s$. This is an {\it a posteriori} consistency check that the interaction between the $D$--brane and the closed strings could be correctly computed in the framework of string theory on Minkowsky space, since the closed strings taking part in this interaction 
``live'' much longer in the flat region $R_2$, complement of $R_1$ in the whole space, rather than in $R_1$. The situation changes when we consider N coinciding $Dp$--branes. An important feature of these objects is that 
parallel $D$--branes do not exert any force on each other and therefore may be set to coincide without expense of energy.
For this system,  the thickness of the curved region $R_1$
around the brane would be proportional to a certain (positive) power of N, and for N large enough, 
it would contain all the physical information on  the interaction between the brane and the closed strings. 
In the limit $N\rightarrow \infty$ $R_1$ would fill the whole space and the interaction of the strings with the branes 
could be described only in the framework of a superstring theory realized on a new curved vacuum defined by the near horizon geometry of the branes. Let us look at this scenario from the low energy supergravity point of view.\par
The system consisting of a large number $N$ of coinciding $D$ branes, is well described by a solution of the low energy supergravity theory since the curvature
in the region $R_1$ is small with respect to $1/{\ell_s}$. 
In general $p$--brane solutions of supergravity are associated with a $p+1$--form potential
$A^{(p+1)}$ to which they  couple and whose integral on their world volume gives their {\it electric} charge. 
(A simple example is that of an elementary particle ($0$--brane) minimally coupled to a vector potential. Its {\it electric} charge is given by the integral of the vector potential along the particle world line.)
If $A^{(p+1)}$
is an elementary field of the theory, then the solution is an {\it elementary} $p$--brane, characterized  by a singular $p$-hypersurface hidden by an event horizon. To an elementary $p$--brane there corresponds a {\it solitonic} $(D-p-4)$--brane coupled to the $(D-p-3)$--form potential {\it dual} to  $A^{(p+1)}$.
This {\it bona fide} solution of the theory is non singular. 
Particular kinds of elementary and solitonic $p$--branes in 
$D$--dimensions can be found
 as solutions of a truncated supergravity model consisting just of the metric, the dilaton and the $(p+1)$--form potential \cite{lup,duff,fre} which are 
described by the following metrics :
\begin{eqnarray}
\mbox{Elementary solution:}\quad &&\nonumber\\
ds^2\,&=&\,\left(1+\frac{k}{r^{\tilde{d}}}\right)^{-\frac{4\tilde{d}}{\Delta(D-2)}}dx^\mu\otimes dx^\nu \eta_{\mu\nu}-\left(1+\frac{k}{r^{\tilde{d}}}\right)^{\frac{4d}{\Delta(D-2)}}dz^p\otimes dz^q \delta_{p q}\nonumber\\
\mbox{Solitonic solution:}\quad &&\nonumber\\
ds^2\,&=&\,\left(1+\frac{k}{r^d}\right)^{-\frac{4d}{\Delta(D-2)}}dx^\mu\otimes dx^\nu \eta_{\mu\nu}-\left(1+\frac{k}{r^d}\right)^{\frac{4
\tilde{d}}{\Delta(D-2)}}dz^p\otimes dz^q \delta_{p q}
\label{branes}
\end{eqnarray}  
The matrix $\eta_{\mu\nu}$ is the metric on a  $(p+1)$--dimensional Minkowsky 
space--time and has the following signature: $\eta_{\mu\nu}={\rm diag}(+,-,\dots,-)$. 
These elementary (solitonic) solutions have a flat world volume parametrized by
the $d=p+1$  ($\tilde{d}=D-p-3$) coordinates $x^\mu$, while $z^p$
denote the coordinates along the directions orthogonal to the world volume
($p=1,\dots, d-p-1$). The parameter $\Delta$ is a characteristic quantity which will have the value $4$
in all the cases we shall deal with. Finally $k$ represents the charge of the solution with respect to the corresponding potential.
The $r=0$ hypersurface represents the {\it event horizon} of the solution and is a coordinate singularity. An important feature of the above solutions 
is to have a residual supersymmetry.\par
In the $11$--dimensional supergravity (low energy limit of the conjectured $M$--theory) there is an $M2$ (elementary) and an $M5$ (solitonic) brane  solution. In type IIB supergravity ($D=10$) of particular interest is the self--dual
(since it is coupled to the R--R self dual $4$--form $A^{(4)}$) $3$--brane solution. Their metric is immediately computed from eqs. (\ref{branes}):
\begin{eqnarray}
\mbox{$D=11$ $M2$--brane: }\quad &&\nonumber\\
ds^2\,&=&\,\left(1+\frac{k_2}{r^6}\right)^{-\frac{2}{3}}dx^\mu\otimes dx^\nu \eta_{\mu\nu}-\left(1+\frac{k_2}{r^6}\right)^{\frac{1}{3}}
\left(d^2r +r^2 d^2\Omega_{7}\right)
\nonumber\\
\mbox{$D=11$ $M5$--brane: }\quad &&\nonumber\\
ds^2\,&=&\,\left(1+\frac{k_5}{r^3}\right)^{-\frac{1}{3}}dx^\mu\otimes dx^\nu \eta_{\mu\nu}-\left(1+\frac{k_5}{r^3}\right)^{\frac{2}{3}}
\left(d^2r +r^2 d^2\Omega_{4}\right)
\nonumber\\
\mbox{$D=10$ $3$--brane: }\quad &&\nonumber\\
ds^2\, &=&\, \left(1+\frac{k_3}{r^4}\right)^{-\frac{1}{2}}dx^\mu\otimes dx^\nu \eta_{\mu\nu}-\left(1+\frac{k_3}{r^4}\right)^{\frac{1}{2}}
\left(d^2r +r^2 d^2\Omega_{5}\right)
\label{M2M5D3}
\end{eqnarray}
The linear extension of the near horizon region $R_1$ may be characterized by
the condition $r\ll R$, $R$ being $k_2^{1/6}$ for the $M2$--brane, 
$k_5^{1/3}$ for the $M5$--brane and $k_3^{1/4}$ for the $3$--brane.
In this region the metrics in eqs. (\ref{M2M5D3}) may be rewritten in the following form:
\begin{eqnarray}
ds^2\,&=&\,\rho^2 dx^\mu\otimes dx^\nu \eta_{\mu\nu}-
(Rw)^2\frac{d\rho^2}{\rho^2}-R^2d\Omega_{(D-p-2)}\nonumber\\
w\,&=&\,\frac{2(D-2)}{\tilde{d}}\,;\,\rho\,=\,\left(\frac{r}{R}\right)^{\frac{1}{w}}
\label{nhorizon}
\end{eqnarray} 
The above metric describes a space--time of the form 
$AdS_{p+2}\times S^{D-p-2}$ (with an abuse of language we shall denote by Anti de Sitter space a particular compactification of Anti de Sitter space, characterized by the condition $\rho>0$). The parameters $k_2,k_5,k_3$ are related to the 
charges these solution have with respect to the potential they couple to. Indeed, in suitable units, they represent the flux of the corresponding 
field strengths
$F^{(4)},F^{(7)},F^{(5)}$ through the spheres $S^7, S^4,S^5$ respectively.
One may interpret the solutions in eqs. (\ref{M2M5D3}) from a microscopic point of view as bound states describing $N$ coinciding $M2$, $M5$, $D3$ branes respectively, each of them carrying a unit of the charge associated with the potential they couple to. In this picture the parameters $k_i$ ($i=2,5,3$)
are proportional to $N$. \par
The solutions (\ref{M2M5D3}) describe a space--time continuum which  starts from a flat Minkowsky geometry 
at infinite distance from the extended object, then acquires a non--vanishing curvature at finite $r$ till it forms a ``throat'' of width $R$ near the horizon at $r=0$, with an anti--de Sitter geometry.\par
Let us spend few more words about the self--dual $3$--brane solution in 
type IIB supergravity. 
If interpreted from the string theory point of view as a bound state of $N$ coinciding $D3$--branes, each carrying  unit of R--R charge, its charge 
with respect to $A^{(4)}$ will be proportional to $k_3=gN\ell_s^4$ ($g$ being the string coupling constant). 
According to the $D3$--brane interpretation of this solution $R\approx \ell_s (gN)^{1/4}$ and therefore in the limit 
$N\rightarrow \infty$ the whole space--time has the form $AdS_5\times S^5$. 
It follows that in this regime a microscopic description of the $D3$--branes 
would be related to superstring theory realized on this new curved background. Indeed spaces of the form $AdS_{p+2}\times M^{D-p-2}$, where $M^{D-p-2}$
is a suitable Einstein manifold, have been shown to be exact solutions of 
superstring or M--theory ($D=10,11$ respectively) \cite{kar} and thus possible vacua for these theories (the space $AdS_5\times S^5$ could be interpreted as a non--perturbative vacuum of superstring theory exhibiting a condensation of $D3$--branes). 
Nevertheless, as previously emphasized, finding the spectrum of physical excitations of a string on a curved background (such as the non--perturbative vacua described above) is a very complicate issue and in general represents an 
unsolved problem (differently from the case of a conformal $\sigma$--model on a Minkowsky background, in a general curved target space indeed it is no more 
 possible to construct the Hilbert space of quantum states as a Fock space built through the action on a vacuum state of creation operators corresponding to free oscillators).
A hint as to the physical content of some of these theories is provided by a powerful duality conjectured by Maldacena almost one year ago. \par
In section $2$ I shall give a pedagogical an tentatively self consistent
 review of this conjecture and 
its formulation as a duality between a {\it singleton} super--conformal 
field theory on the boundary of Anti de Sitter space and superstring or
M theory in the bulk. In section $3$ I shall focus on the particular case of the $M2$--brane and discuss some recent results 
in the analysis of the dynamics of branes in Anti de Sitter space.   
\section{$AdS/CFT$ Duality}
Let us focus for the moment on a $10$--dimensional space--time in which the physics is described by type IIB superstring theory in presence of $D3$--branes.
The main idea is that, as the number $N$ of coinciding $D3$--branes
increases ($N\gg 1$), the physics is described by a new non--perturbative 
vacuum which is no more the flat Minkowsky one on which the superstring theory is perturbatively defined but has the geometry of $(AdS_5\times K^5)_N$ in which the $N$ $D3$--branes are suitably embedded (the subscript ``$N$'' reminds 
that the background  space--time is the near horizon geometry of the $N$ branes and therefore its ``radius'' is proportional to $N^{1/4}$) . 
It is reasonable to think that the 
 effective low energy theory around this vacuum has $D3$--branes as fundamental objects instead of strings and describes fluctuations of these extended objects, around their static configuration  described by eq. (\ref{M2M5D3}), which have energy much lower than the string scale ($1/\alpha^\prime$). This theory is the effective field theory on the world volume on the $N$ coinciding branes in the limit $\alpha^\prime\rightarrow 0$
(in this limit the interaction of the branes with the bulk through emission of closed strings is suppressed).
 Intuitively, in much the same way as a field theory of 
elementary particles and of solitons can be viewed as effective descriptions of a same quantum theory on two 
different vacua, the theory of superstrings and of branes may be considered as ``dual'' to each other, i.e. the theory defined on the world volume of the $N$ coinciding branes (in which the branes are the fundamental objects) embedded in their near horizon 
geometry, for small fluctuations, could be viewed as the effective low energy realization of the quantum 
theory of strings on an $(AdS_5\times K^5)_N$ vacuum. 
This is the basic idea which has been formalized by Maldacena in his powerful duality conjecture:\par
{\it Type IIB superstring theory realized on $(AdS_5\times S^5)_N$ and the low 
energy effective field theory on the world volume of the $N$ coinciding
$D3$--branes are the same quantum theory.} \par
The same kind of duality was originally conjectured also between 
$M$--theory (whose physical content is not known so far) on 
$(AdS_{p+2}\times S^{9-p})_N$, describing the near horizon geometry of $N$
$Mp$--branes for $p=2,5$, and the low energy effective field theory on the world volume of these branes. In this case the low energy condition on the world volume theory is expressed in terms of the only length scale of the $11$ dimensional supergravity theory, which is the Plank scale $\ell_p$:  
$\ell_p\rightarrow 0$.
This duality is also referred to as {\it holographic} correspondence
since it states that the quantum dynamics of fields in a $(p+2)$--dimensional space--time
($AdS_{(p+2)}$) is encoded in a theory defined on a $(p+1)$
dimensional subspace. \par
In the original formulation of the conjecture the inner compact space is a 
$(D-p-2)$--dimensional sphere $S^{(D-p-2)}$.
Eventually it was suggested that the duality could be formulated on more general spaces in which the compact Einstein space $K^{(D-p-2)}$ is an homogeneous space of the form $G/H$ \cite{gh}. These spaces are maximally symmetric solutions of the supergravity theory and near horizon geometry of certain $p$--brane solutions of the same theory. Differently form the case in which $K=Sphere$, solutions of the form  $AdS_{(p+2)}\times (G/H)$
are in general not maximally supersymmetric, i.e. they preserve less supersymmetry than the original theory.
Maldacena's conjecture has been recently further extended to spacetime geometries of the form $AdS\times K$ 
in which $K$ is an even more general Einstein manifold related to the so called Sasaki monifolds \cite{sasaki}.\par
A special role in this duality is played by the Anti de Sitter space--time
 representing the non--compact  factor  of this vacuum. Superstring theory
(supergravity) realized on $AdS_{(p+2)}\times K^{(D-p-2)}$ is {\it locally} 
invariant under the general coordinate transformations generated by the 
space time isometry group:
\begin{equation}
{\cal G}\,=\,{\cal I}som \left(AdS_{(p+2)}\times K^{(D-p-2)} \right)\,=\, SO(2,p+1)\otimes
{\cal I}som \left( K^{(D-p-2)} \right)
\label{isom}
\end{equation} 
Since we are dealing with a locally supersymmetric theory, we may include 
supersymmetry transformations in the previous statement and say that the quantum theory in the bulk is locally invariant with respect to the superextension 
${\cal SG}$ of the isometry group in (\ref{isom}), whose structure will be 
 described in more detail in next section. However it is useful to write
the explicit form of ${\cal SG}$ in the following way:
\begin{eqnarray}
{\cal I}som \left( K^{(D-p-2)}\right)&&\rightarrow SO(N)\oplus {\cal K}^\prime
\nonumber\\  
{\cal SG}\,&=&\, {\cal SC}\oplus {\cal K}^\prime\nonumber\\
{\cal SC}\,&=&\,\cases{Osp(4/{\cal N}) & \mbox{$M2$--brane}\cr
Osp(6,2/{\cal N}) & \mbox{$M5$--brane}\cr SU(2,2/{\cal N} ) & \mbox{$D3$--brane}}
\label{sg}
\end{eqnarray}
where ${\cal N}\times$(dimension of the spinors in $(p+2)$ dimensions) 
gives the number of supercharges preserved by the background solution
(${\cal N}$ is the number of {\it Killing spinors} of the supergravity solution and depends on the compact manifold $K$). The group $ {\cal SC}$ is the supersymmetric extension of the group $SO(2,p+1)$, that is it consists of the 
 $SO(2,p+1)$ bosonic generators plus a number ${\cal N}$ of fermionic 
generators. An important property of the groups $SO(2,p+1)$ and $ {\cal SC}$ is that they act  respectively 
as the {\it conformal} and {\it super conformal} groups in $p+1$ dimensions.\par 
One of the features of an Anti de Sitter space is to have a boundary.
This boundary $\partial  AdS_{(p+2)}$ is a $(p+1)$ dimensional 
locus of points having the property of being stable with respect to the action 
 of the $AdS_{(p+2)}$ isometry group $SO(2,p+1)$ (or the superisometry group ${\cal SC}$
of the Anti de Sitter superspace, in the framework of a supersymmetric theory).\par
Parametrizing $AdS_{(p+2)}$ by means of the coordinates $(\rho, x^\mu)$ 
in terms of which the metric is written in the form (\ref{nhorizon}), the boundary may be characterized as the $(p+1)$--dimensional Minkowsky space $M_{(p+1)}$ 
spanned by the coordinates $x^\mu$ in the limit $\rho\rightarrow 0$ plus a point at infinity $\rho\rightarrow \infty$. Therefore the Minkowsky part of 
$\partial  AdS_{(p+2)}$ ($M_{(p+1)}$) 
coincides with the world volume of the $N$ overlapping $p$--branes, which is thus stable under the action of the $SO(2,p+1)$
isometry group.\par The effective low energy theory on the world volume 
$M_{(p+1)}$ of the $N$ $p$ branes is obtained by expanding the Born--Infeld 
action $S_{BI}$ on $M_{(p+1)}$ for small oscillations of the branes around the static position at the boundary of  $AdS_{(p+2)}$. The action  $S_{BI}$
describes a $\sigma$--model on $ M_{(p+1)}$ in which the local fields are 
the coordinates $X^M(\xi)=(\rho(\xi), x^\mu(\xi);y^m(\xi))$ of the target 
space--time 
$AdS_{(p+2)}\times K^{(D-p-2)}$, locally depending on the $p+1$ world volume coordinates $\xi^a$. 
The main  feature of  $S_{BI}$ is that it is expressed as the integral 
over  $ M_{(p+1)}$ of the invariant volume element and thus the isometries of the target space are {\it global} symmetries of the theory. In particular, since the Anti de Sitter isometry group $SO(2,p+1)$ acts on $ M_{(p+1)}$ as a conformal group, the theory on the world volume of the $p$--branes on the boundary 
is a {\it conformal field theory} (CFT). If  supersymmetries are taken into account, the theory is expected  to be globally invariant under 
the whole superisometry group ${\cal SG}$, which contains the superconformal group ${\cal SC}$, and thus to be a {\it super conformal} field theory
(SCFT).\par 
The conjectured
duality is therefore between superstring ($M$) theory on  $AdS_{(p+2)}\times K^{(D-p-2)}$ and a SCFT on the boundary $\partial  AdS_{(p+2)}$. \par
This conjecture has been subsequently made more precise by Witten 
\cite{witten} who defined a precise correspondence between quantum states
$\Phi (x)$ of the theory in the bulk and superconformal operators 
${\cal O}(\xi)$ on the boundary. In particular, according to this picture,
the operators  ${\cal O}(\xi)$ are interpreted as vertex operators defining the emission of the corresponding state $\Phi (x)$ from the membrane in  the bulk.
At the heart of this correspondence is the consideration that either 
the states $\Phi (x)$ or the operators ${\cal O}(\xi)$ must transform in representations of the same group ${\cal G}$ defined in eq. (\ref{isom}) or, more in general, of its superextension ${\cal SG}$. Witten retrieved also a 
relation between the masses of the states in the bulk and the conformal dimensions of the operators on the boundary. \par
As I previously emphasized the whole spectrum of the quantum theory in the bulk is not known, nevertheless the low energy part of it consists of the Kaluza--Klein states  
of $D$--dimensional supergravity compactified on $AdS_{(p+2)}\times K^{(D-p-2)}$, which have been extensively studied in the eighties \cite{kaluza}.
These states are in one to one correspondence with the harmonic functions on the 
compact space $ K^{(D-p-2)}$ and their mass is of order $m_{KK}\approx 1/R$.
They sit in {\it short} supermultiplets of the supergroup ${\cal SG}$ since their masses are related to quantities characteristic of the isometry algebra of 
 $ K^{(D-p-2)}$ and this property protects them, for large enough supersymmetry
, from quantum corrections. In the previously outlined correspondence, 
the Kaluza--Klein states naturally 
correspond to conformal operators sitting in short representations of the superconformal group ${\cal SC}$ acting on the boundary. They are the {\it chiral} operators.
Besides the Kaluza--Klein states there are the string or $M$ theory states
which sit on {\it long} supermultiplets of ${\cal SG}$. Their mass is of order
$m\approx 1/\ell$ ( $\ell$ being $\ell_s$ or $\ell_p$ in the 
superstring or $M$--theory case respectively) and are not protected from quantum corrections. These  states would naturally correspond to {\it non chiral} operators in the SCFT on the boundary. \par
 Let us recall that the ``radius'' of the Anti de Sitter space $R$ depends on both $\ell$ and $N$ in the following way:
\begin{eqnarray}
&& R\approx \ell N^\beta
\end{eqnarray}
where $\beta$ is $1/4$, $1/6$, $1/3$ for the $D3$, $M2$, $M5$ brane respectively. If we send, together with $\ell \rightarrow 0$ (low energy limit on 
world volume theory) $N$ to infinity in such a way as to keep $R$ fixed, the string or $M$ theory modes will decouple since their mass will grow as $N^\beta$.
In this limit the duality can be restated as a mapping between a SCFT on the boundary and tree--level Kaluza--Klein supergravity in the bulk. Since much is known about the latter, Maldacena's duality has so far 
received several checks from the study of the large $N$ limit of 
certain SCFT (especially in the case of  the low energy SCFT on 
the world volume of $N$ coinciding $D3$--branes which is known to be a super 
conformal $SU(N)$ Yang--Mills theory. In this framework stringent check of the duality was carried out for instance in \cite{van}).\par
In the correspondence described above between states of the theory in the bulk and conformal operators on the boundary belonging to the same representation of the superconformal group, there is a piece missing. Indeed, among the unitary irreducible representations, besides the long and short supermultiplet, 
the superconformal group ${\cal SC}$ admits {\it ultrashort} representations
({\it super--singletons}) (in a non--supersymmetric framework, the ultrashort representations of the AdS isometry group $SO(2,p+1)$ are called {\it singletons} and were first found by Dirac \cite{Dirac}). Supersingletons may be characterized as 
those states which saturate the lowest bound on the energy in order for the representation to be unitary.
States in these multiplets  don't have a representation in terms of  
local fields in the bulk since they can be gauged away everywhere, except on the boundary of the Anti de Sitter space, where they admit a field representation
. Super--singletons are therefore the natural candidates 
for describing the elementary fields of the SCFT on the boundary. They don't have a counterpart in the theory defined on the bulk. Superconformal field theories of singletons on $\partial AdS$ have been studied extensively in the eighties \cite{singleton}.
An important property of these ultrashort representations is that all 
the other unitary irreducible representations (UIR) of ${\cal SC}$ (in which 
for instance all the fields in the bulk transform) may be built from 
tensor products of two or more singleton states (composite states).
 This lead, during the early stages of the research on  singleton field 
theories, to the hypothesis that all the excitations of the quantum theory in the bulk could be expressed as composite 
states of elementary super--singletons living in the boundary (see \cite{duff} and references therein) which was a first kind of {\it holographic } correspondence.\par
Already in these early years moreover the super--singleton SCFT
was suggested (Nicolai et al. \cite{nicolai}) to describe the dynamics of physical objects embedded on the boundary of $AdS_{(p+2)}$, namely $p$--branes. However this correspondence was never proven rigorously and there are two main differences between the super--singleton SCFT studied in the eighties and those which are expected
 to describe one side of Maldacena duality. The former were defined 
on a boundary of $AdS_{(p+2)}$ with topology $S^p\times S^1$, they were 
(for large enough supersymmetry) free field theories with a mass term  and were related to 
the dynamics of a single {\it spherical } $p$--brane  embedded in $\partial AdS_{(p+2)}$. On the other hand SCFTs relevant to Maldacena conjecture 
(as previously pointed out) are defined on a boundary
$\partial AdS_{(p+2)}$ which is given the topology of a $(p+1)$--dimensional  
Minkowsky space (plus a point at infinity), are interacting (since they 
 have a non--abelian $SU(N)$ gauge invariance) massless field theories
and are expected to describe the dynamics of $N$ coinciding $p$--branes
set on $\partial AdS_{(p+2)}$. \par
Nevertheless, the characterization of the SCFT entering Maldacena duality
as a super--singleton theory allowed for a group theoretical analysis 
of the conjecture \cite{andria}.\par
There is however an important feature of the super--singleton theory which is 
in general not fixed by its invariance with respect to ${\cal SC}$. Indeed 
we recall that this theory is globally invariant with respect to the whole 
superisometry group ${\cal SG}$ in eq. (\ref{sg}) which contains, besides ${\cal SC}$, a bosonic subgroup ${\cal K}^\prime$. This group acts as a {\it flavour} group on the singleton theory  and the flavour representation of the singletons 
with respect to ${\cal K}^\prime$ is independent of their superconformal transformation properties and can in general be retrieved only from the $p$--brane dynamics on the boundary. In the maximally supersymmetric case $K^{(D-p-2)}=S^{(D-p-2)}$ the whole ${\cal I}som \left(S^{(D-p-2)}\right)=SO(D-p-1)$ enters 
the superconformal group ${\cal SC}$ in eqs. (\ref{sg}) and the group 
${\cal K}^\prime$ is trivial, therefore all the transformation properties 
of the singletons with respect to ${\cal SG}$ are characterized  by the 
${\cal SC}$ quantum numbers. Verifying Maldacena conjecture in the most general case requires the knowledge of the flavour representation of the 
singletons with respect to ${\cal K}^\prime$. Indeed since all the 
conformal operators of the theory  are constructed as composite structures
of the singleton fields, their  ${\cal K}^\prime$ quantum numbers 
will depend on the singleton flavour representation. Verifying the matching
between the ${\cal K}^\prime$ quantum numbers of the boundary operators with
those of 
 the bulk states would be a new kind of check of Maldacena conjecture.   
To this end it is crucial to develop a procedure for retrieving the 
supersingleton action from the quantization of a 
$N$ coinciding $p$--branes set on the boundary of $ AdS_{(p+2)}$. For $N>1$
a classical action on the world volume of the overlapping branes 
would be a non--abelian Born--Infeld action, which is not known at present.
However, if our aim is just to determine the ${\cal K}^\prime$ flavour representation of the singletons, this information could be inferred from the study of a single $p$--brane on the boundary. Indeed one may consider a system of $N+1$
parallel $p$--branes, of which the first $N\gg 1$ are coinciding and set on the boundary of $AdS_{(p+2)}\times  K^{(D-p-2)}$ while the remaining one 
({\it probe} brane) is set at a distance $r$ from them. In principle then one could quantize (for small fluctuations) the world volume Born--Infeld action on the probe brane in the limit in which the latter is sent to coincide with the other $N$ 
($r\rightarrow 0$) (superconformal limit) and retrieve an the superconformal 
action of free (abelian) massless singletons, together with their 
${\cal K}^\prime$ representation.\par Quantizing for small oscillations 
the B--I action on the world volume of a brane embedded in a curved space 
is a difficult problem because of technical difficulties related to 
the gauge fixing of the local invariances of the theory (in particular 
the fermionic $\kappa$ symmetry needed in order for the world volume theory to
 be supersymmetric). A general method for retrieving the supersingleton action
from the world volume theory of a probe brane on the boundary of an Anti de Sitter space was defined in \cite{osp}, and applied to the case of an $M2$--brane
on the boundary of $AdS_4\times S^7$. This method relied on a 
particular parametrization of the Anti de Sitter superspace defined by using 
a {\it supersolvable} subalgebra of the superisometry group. The supersolvable parametrization is the super extension of the {\it solvable} parametrization
of the Anti de Sitter space defined in \cite{gh}, which is related to the coordinates $(\rho, x^\mu)$ in terms of which the metric has the expression (\ref{nhorizon}). The main goal of the supersolvable parametrization is to allow 
to write the B--I action which is $\kappa$--symmetry fixed from the very beginning, that is which depends only on the physical fermionic fields. This $\kappa$--gauge fixing turns out to be equivalent to fixing the  {\it Killing spinor} gauge defined in \cite{kall}.\par Solvable and Supersolvable parametrizations defined in \cite{gh} and \cite{osp} revealed to be a powerful algebraic tool
for describing brane dynamics in Anti de Sitter spaces. They have been applied 
eventually to the $D3$ brane in $AdS_5\times S^5$ \cite{pesando}.\par
The final goal however is to apply this method to a background
of the form $AdS_{(p+2)}\times K^{(D-p-2)}$ where $K$ is an homogeneous 
manifold of the form $G/H$ or related to a more general Sasaki manifold (${\cal N}=2$ Killing spinors).
In these cases indeed, as previously emphasized, the $AdS/CFT$
correspondence from a group theoretical point of view becomes non--trivial 
because the ${\cal K}^\prime$ flavour representation of the singletons 
enters the game on the SCFT side, as an input independent of the 
superconformal invariance of the theory. This analysis is still work in progress.
\section{Supermembrane in Anti de Sitter space.}
In the present section I shall recall the main facts about Anti de Sitter 
spaces and review some recent results in the study of the
$M2$--brane dynamics in  $AdS_4\times S^7$.\par
An $n$--dimensional Anti de Sitter space--time may be described as 
an hypersurface embedded in $\IR^{2,n-1}$. Let $X^\Lambda=(X^{0},X^{\bar{0}},
X^1,\dots, X^{n-1})$ denote a coordinate basis for $\IR^{2,n-1}$ with respect to which the flat metric has the form: $\eta_{ \Lambda\Sigma}={\rm diag}(+,+,-,\dots,-)$. $AdS_n$ is defined as the following locus of points:
\begin{eqnarray}
\eta_{ \Lambda\Sigma}X^\Lambda X^\Sigma\,&=&\, 1
\label{hyper}
\end{eqnarray}
(for simplicity we have set the ``radius'' of $AdS_n$ equal to $1$).
This condition is invariant with respect to the group $SO(2,n-1)$ acting on $X^\Lambda$, which is the isometry group of $AdS_n$. 
Defining the light cone coordinates $X^{\pm}=X^{\bar{0}}\pm
X^{n-1}$ and the coordinates $x^\mu$ by setting $X^\mu=X^+x^\mu $ for $\mu=0,1,\dots,n-2$, the condition (\ref{hyper}) may be rewritten in the following way:
\begin{eqnarray}
X^+X^-+(X^{+})^2\eta_{ \mu\nu}x^\mu x^\nu \,&=&\, 1
\label{hyper2}
\end{eqnarray}
Let us restrict ourselves to the branch $X^+\ge 0$. In the limit $X^+\gg 1$ the above condition reduces to:
\begin{eqnarray}
X^{+}\eta_{ \mu\nu}x^\mu x^\nu \,&=&\, -X^-
\end{eqnarray}
which defines an $(n-1)$--dimensional Minkowsky space $M_{(n-1)}$ spanned by the coordinates $x^\mu$, which is stable with respect to a scaling $X^\Lambda\rightarrow \lambda X^\Lambda$. Its isometry group is $ISO(1,n-2)\subset SO(2,n-1)$. Moreover if we add to it the  point P at $X^+=0$
defining $\tilde{M}_{(n-1)}=M_{(n-1)} \cup P(X^+=0)$, this 
compact locus of points defines the boundary of $AdS_n$ which is stable with respect to the whole group $SO(2,n-1)$. Setting $\rho=X^+$ and computing the induced metric on the hypersurface (\ref{hyper2}) in terms of the coordinates
$(\rho,x^\mu)$  one obtains a generalization of the 
Bertotti--Robinson metric:
\begin{eqnarray}
ds^2=\rho^2 dx^\mu \otimes dx^\nu \eta_{\mu\nu} -\frac{d\rho^2}{\rho^2}
\label{BR}
\end{eqnarray} 
The Anti de Sitter space may be alternatively written as an homogeneous manifold in the following way:
\begin{equation}
AdS_n=\frac{SO(2,n-1)}{SO(1,n-1)}
\label{coset}
\end{equation}
In order to compute  the non--linear action of the isometry group 
$SO(2,n-1)$ on the coordinates $(\rho, x^\mu)$, we shall characterize them 
as coordinates in the  {\it solvable} representation of $AdS_n$
\cite{gh}.
Using the formalism of coset manifolds, one can associate with each point Q 
in $AdS_n$ a {\it coset representative} $L(Q)$ so that the non linear action 
of the isometry group $SO(2,n-1)$ on the point Q may be represented in the following way:
\begin{eqnarray}
\forall g\in SO(2,n-1)&&\,\,\, g\cdot L(Q)\,=\, L(Q^\prime(g,Q))\cdot h(g,Q)\nonumber\\
h(g,Q)\in SO(1,n-1)
\label{g}
\end{eqnarray} 
In order to define a parametrization of the manifold, or equivalently  $L(Q)$,
one may use the following algebra decomposition:
\begin{eqnarray}
SO(2,n-1)\,&=&\, SO(1,n-1)\oplus Solv
\label{iwa}
\end{eqnarray} 
where $Solv$ is an $n$--dimensional {\it solvable}\footnote{A solvable Lie algebra $Solv$ is a Lie algebra whose $k^{th}$ Lie derivative ${\cal D}^k(Solv)$ vanishes for some finite $k$. The $k^{th}$ Lie derivative may be defined by induction in $k$:
${\cal D}^{k+1}(Solv)=\left[{\cal D}^{k}(Solv),{\cal D}^{k}(Solv)\right]$; 
${\cal D}^{1}(Solv)=\left[Solv,Solv\right]$ } subalgebra of $SO(2,n-1)$.
As a consequence of eq. (\ref{iwa}) the Anti de Sitter space is isomorphic to a group manifold generated by $Solv$ and may be globally identified 
with it. In this description  the local coordinates on 
$AdS_n$ are therefore parameters of $Solv$. The $Solv$ algebra is constructed by considering the only non--compact Cartan generator $D$ in the coset (\ref{coset})
and diagonalizing its adjoint action on the algebra $SO(2,n-1)$. This defines a {\it grading} on $SO(2,n-1)$ which is decomposed into the direct sum of 
eigenspaces of $Adj_D$, namely $g_{(0)},g_{(\pm1)}$ corresponding to the eigenvalues $0$, $\pm1$ respectively. $Solv$ can be constructed as $\{D\}\oplus
g_{(-1)}$, where $g_{(0)}=\{D\}\oplus SO(1,n-2)$ while $g_{(-1)}=\{T_\mu\}$ consists of       
the $n-1$ {\it shift operators} corresponding to all the roots having negative 
value on $D$. The subalgebra $g_{(-1)}$ is an abelian subalgebra of $SO(2,n-1)$ since there is no generator with grading $-2$ with respect to $Adj_D$.
Associating with the generator $D$ the parameter ${\rm log}(\rho)$ and 
with the generators $T_\mu$ the parameters $x^\mu$, the coset representative 
can be defined as an element of the solvable group 
${\rm Exp}(Solv)$ as follows:
\begin{equation}
L(\rho,x^\mu)\,=\,{\rm Exp}\left({\rm log}(\rho) D+ \sum_{\mu=0}^{n-1}x^\mu
T_\mu\right)
\label{cosetrep} 
\end{equation} 
Constructing the vielbeins in this parametrization and then computing the metric on the manifold, one obtains the expression in eq. (\ref{BR}).
From eqs. (\ref{cosetrep}) and (\ref{g}) one may therefore 
deduce the action of $SO(2,n-1)$ on the coordinates $(\rho, x^\mu)$. \par 
In this parametrization Anti de Sitter space is represented, roughly speaking, by means of Minkowsky 
``slices'' $M_{(n-1)}$ spanned by the coordinate $x^\mu$ and fibrated 
along the transverse coordinate $\rho \in \IR^+$.One may define an 
alternative topology on $\partial AdS_{n}$, by constructing it as the 
limit of $M_{(n-1)}$
as $\rho\rightarrow 0$ plus a point at $\rho\rightarrow \infty$ \cite{osp}. 
Thinking of $AdS_n$ as the near horizon geometry on $N$ coinciding 
$(n-2)$ branes set on the $\rho=0$ chart of the boundary,  the 
{\it probe} brane introduced in last section, can be defined as a brane whose wold volume coincides with the $M_{(n-1)}$
Minkowsky ``slice'' at a distance $\rho$ from the remaining $N$ branes.\par
It is possible to check that the action of the group $SO(2,n-1)$ on 
a generic hypersurface $M_{(n-1)}$ at $\rho\neq 0$ closes a ``soft'' conformal algebra, that is a conformal algebra whose structure constants depend on the coordinate $\rho$ ( the {\it broken conformal transformations} of \cite{malda}). 
In the limit $\rho\rightarrow 0$ (on the boundary)
the dependence on $\rho$ of the conformal algebra drops out and the group
$SO(2,n-1)$ acts on $M_{(n-1)}$ as the usual conformal group. Now we can interpret the generators of $SO(2,n-1)$ described above, from the world volume theory point of view,  in the following way:
\begin{itemize}
\item $D$ the generator of the {\it dilations}
\item $T_\mu$ the $n-1$ {\it translations} on $M_{(n-1)}$
\item $K_\mu$ in $g_{(+1)}$ as the $n-1$ {\it special conformal transformations}
\item $SO(1,n-2)=g_{(0)}/\{D\}$ the {\it Lorentz transformations} on $M_{(n-1)}$
\end{itemize}
In what follows I shall outline, without entering the details of the 
calculations, the main conceptual steps in the procedure 
introduced in \cite{osp} for computing the supersingleton action from the 
quantum fluctuations of a probe $p$--brane around  the boundary 
of an Anti de Sitter space. 
Let us consider for simplicity the special case of an $M2$ probe brane
in $AdS_4\times S^7$. As previously pointed out this space--time is a 
maximally supersymmetric solution of the $11$--dimensional supergravity.
 With it we may associate a superspace $AdS^{(4/8)}\otimes S^7$ 
spanned by $10$
bosonic coorsinates $X^M$ ($M=0,\dots, 9$) and $32$ fermionic coordinates 
$\Theta^A_\alpha$ where $A=1,\dots, {\cal N}=8$ and $\alpha=1,\dots,4$
indicizes  a $4$--dimensional spinor. $X^M$ consist in the four coordinates
$(\rho, x^\mu)$ of $AdS_4$ and the seven coordinates $y^m$ of $S^7$.
The superspace has the form:
\begin{eqnarray}
AdS^{(4/8)}\otimes S^7\,&=&\, \frac{Osp(4/8)}{SO(1,3)\times SO(8)}\otimes S^7
\label{sspace}
\end{eqnarray}
The theory on the world volume $M_{(3)}$ of the $M2$ probe brane is described 
by a Born--Infeld action $S_{BI}$
which, as explained in the previous section, is a $\sigma$--model defined on 
 $M_{(3)}$ having the background superspace
$AdS^{(4/8)}\otimes S^7$ as target space. The local fields in this action are therefore  
$X^M(\xi)$ (bosons) and $\Theta^A_\alpha(\xi)$ (fermions), $\xi^a$ being the $3$ world volume coordinates. Using the conventions of \cite{osp} $S_{BI}$ may be written 
(in suitable units) in the following form:
\begin{eqnarray}
S_{BI}\,&=&\, 2\int_{M_{(3)}}\sqrt{-{\rm det}(h_{ab})}+(4!)^2\int_{M_{(3)}}
 A^{(3)}
\label{biaction}
\end{eqnarray}
where $h_{ab}(X^M(\xi),\Theta^A_\alpha(\xi))$ is the induced metric on the world volume and the second term is the Wess--Zumino term describing the minimal coupling of the brane with the $3$--form of the $11$--dimensional supergravity.\par Let us analyze the symmetries of this action. As previously stated, its {\it
global} symmetries are the target space superisometries which close the group
${\cal SG}=Osp(4/8)$. Its {\it local} symmetries are, on the other hand, 
 the {\it diffeomorphisms}
on the world volume ($\xi\rightarrow \xi^\prime(\xi)$) and the fermionic
$\kappa$--symmetry. The former implies that the only physical degrees of 
freedom are the oscillations of the membrane in the directions perpendicular to its world volume ($8$ bosonic d.o.f.). The latter is a symmetry of the action with respect to local transformations acting only on 
half the components of the fermion fields. These components are defined by a 
projector of the form ${\cal P}^+=(1+\Gamma)/2$, where $\Gamma$ is a
$4\times 4$ matrix depending on the background fields. The invariance
of $S_{BI}$ with respect to this local transformation requires the background fields to be a solution of the $11$ dimensional supergravity, which is indeed the case.The consequence of $\kappa$ symmetry is that the $\Theta^+={\cal P}^+\Theta$ components of the fermionic fields can be gauged away, leaving 
$16$ physical fermionic degrees of freedom $\Theta^-$ which match 
{\it on--shell} the $8$ bosonic ones. 
The super--singletons indeed belong to  $8$ supermultiplets
consisting of a boson and a Majorana fermion each. These fields will
 be retrieved as quantum fluctuations around a suitable solution of the $8$ 
fermionic and $8$ bosonic physical degrees of freedom of the brane.
The existence of $\kappa$ symmetry is therefore required 
in order for the theory on the world volume to be on--shell supersymmetric.\par
A particular solution of the theory is the static one:
\begin{eqnarray}
x^\mu(\xi)\, &=&\, \xi^a \,\, \mu,\, a\,=\, 0,1,2\nonumber\\
\rho(\xi)\,&\equiv&\, \bar{\rho}\,=\, const\nonumber\\
\partial_a y^m\, &=&\,0\nonumber\\
 \Theta^A_\alpha\, &=&\,0
\label{static}
\end{eqnarray} 
In order to retrieve the supersingleton action the strategy is to fix the local symmetries of the theory and expand the action (\ref{biaction}) for small fluctuations of the physical d.o.f. around the solution (\ref{static}) and then 
taking the boundary limit $\bar{\rho}\rightarrow 0$ where the full 
superconformal symmetry is restored. The local diffeomorphisms are fixed by 
fixing the coordinates $x^\mu$ parallel to the brane to the solution (\ref{static}). Fixing the $\kappa$ symmetry is a more complicate issue. In principle 
one should compute the components $\Theta^\pm$ on the static solution, set
the unphysical components $\Theta^+$ to zero in the action and allow the physical ones to fluctuate. This is difficult to implement on the action  (\ref{biaction}) initially expressed in terms of the whole spinors $\Theta_\alpha^A$, since on the Anti de Sitter background, this dependence is rather complicate
and involves higher powers of the $\Theta$ fields.\par This problem has been circumvented in \cite{osp} by using the supersolvable parametrization of the superspace (\ref{sspace}) which consists in  
redefining the target space of our $\sigma$
model to be a subspace of the whole superspace which is a supermanifold generated by a {\it supersolvable} algebra ($SSolv$) (times of course the sphere $S^7$spanned by $y^m$).\footnote{The definition of 
a supersolvable Lie algebra is the same as that of a solvable algebra 
but with the supercommutator substituted to the commutator} We start from a decomposition analogous to that in (\ref{iwa}):
\begin{eqnarray}
Osp(4/8)\,&=&\, \left[SO(1,3)\oplus SO(8)\oplus {\cal Q}\right]\oplus Ssolv
\end{eqnarray}
The above decomposition is obtained by performing a grading of the superalgebra with respect to $Adj_D$:
\begin{eqnarray}
Osp(4/8)\,&&\rightarrow g_{(-1)}\oplus sg_{(-1/2)}\oplus g_{(0)}\oplus 
sg_{(1/2)}\oplus g_{(1)}\nonumber\\
{\cal Q}\,&=&\,sg_{(1/2)}\nonumber\\
{\cal S}\,&=&\,sg_{(-1/2)}\nonumber\\
 SO(1,3)\oplus SO(8) \oplus \{D\}\,&=&\,g_{(0)}\nonumber\\
SSolv\,&=&\,\{D\}\oplus g_{(-1)}\oplus sg_{(-1/2)}\,=\, Solv\oplus sg_{(-1/2)}
\end{eqnarray}
the spaces $g_{(\pm 1)}$ are the same as those defined previously for the 
Anti de Sitter space. The subspaces $sg_{(\pm 1/2)}$ are the  eigenspaces 
with of the fermionic generators corresponding to the eigenvalues $\pm 1/2$ of 
$Adj_D$: on the world volume theory the ${\cal S}$ operators ($16$ components) generate the {\it special superconformal
transformations}, while  the ${\cal Q}$ operators ($16$ components) generate the {\it supersymmetry transformations}. Differently from the solvable representation of Anti de Sitter space, 
the supergroup ${\rm Exp} (SSolv)$ does not coincide with the original superspace since the generators ${\cal Q}$ are modded out. Nevertheless it can be shown that $Adj_D$ on the fermionic generators is represented by an operator $\tilde{\Gamma}$ which coincides with the $\kappa$ symmetry operator $\Gamma$ on the static solution (\ref{static}) (it is expressed as
 the product of the gamma matrices along the directions of the world volume). 
In other words, the $\kappa$--symmetry projectors ${\cal P}^\pm$ computed on the solution (\ref{static}) coincide with the projectors of the supersymmetry generators into the eigenspaces $sg_{(\pm 1/2)}$ respectively.
Therefore the fermionic coordinates parametrizing 
$SSolv$ are already the $\kappa$ gauge fixed ones. 
In this parametrization  indeed the coordinates are 
$(\rho, x^\mu, y^m ; \Theta^-)$. The B--I action defined 
on this supersolvable target space therefore is $\kappa$ gauge fixed from the very beginning and much simpler to compute. \par 
Expanding this action around the static solution for small fluctuations of the physical fields, rescaling the latter by suitable powers of $\bar{\rho}$, taking the order $\alpha^{\prime 0}$ of the action and sending 
$\bar{\rho}\rightarrow 0$ it was possible to retrieve the supersingleton action
on the boundary as a free superconformal field theory describing $8$ massless
bosons and $8$ massless Majorana fermions.
\section{Conclusions}
The aim of the present talk is on one hand to give a tentatively self--contained and hopefully elementary introduction to Maldacena's conjecture
and on the other hand to frame within a discussion on the intense research devoted to verify this conjecture and to possibly extend it, 
some recent results in the study of brane dynamics in certain Anti de Sitter spaces.  I emphasized how some aspects of the super--singleton
CFT on one side of this duality could be inferred only from the dynamics of $p$--branes on the boundary of a $AdS_{(p+2)}\times K^{(D-p-2)}$
space--time for a general compact Einstein manifold $K$. The knowledge of these aspects would provide  the basis for a new stringent check of 
the conjecture. So far a method for constructing the super--singleton action on the world volume of a probe brane 
at the boundary of  $AdS_{(p+2)}\times S^{(D-p-2)}$ has been defined by means of a {\it normal coordinate expansion} (small quantum fluctuations)
of the B--I action on the probe brane around the static configuration in which the latter is set to lie on the boundary of the Anti  de Sitter space.
An extension of this method to more general internal spaces $K$ is still work in progress.

\end{document}